\newcommand{\Pl}{\partial}
\newcommand{\ts}{\textstyle}
\newcommand{\bee}{\begin{equation}}
\newcommand{\ene}{\end{equation}}
\newcommand{\beea}{\begin{eqnarray}}
\newcommand{\enea}{\end{eqnarray}}

\newcommand{\fpar}[2]{\frac{{\ts \Pl \/ #1}}{{\ts \Pl \/ #2}}}
\newcommand{\nder}[3]{\frac{{\ts d^{#1} \/ #2}}{{\ts d \/ #3^{#1}}}}

\documentclass[pre]{revtex4}
\usepackage{dcolumn}
\usepackage{graphicx}
\usepackage{latexsym}
\usepackage{amsmath}
\usepackage{amssymb}
\usepackage{amsfonts}
\usepackage{amsthm}
\begin{document}
 \title{ Shear Flow instability in a   strongly coupled dusty plasma}
 \author{D. Banerjee, M. S. Janaki and N. Chakrabarti}
 \affiliation{ Saha Institute of Nuclear Physics,
 1/AF Bidhannagar Calcutta - 700 064, India.}
%---------------------------------------------------------------------------------------
\begin{abstract}
Linear stability analysis of strongly coupled incompressible dusty plasma  in presence of  shear flow  has been carried out using
Generalized Hydrodynamical (GH) model.  With the proper Galilean invariant  GH model, a nonlocal eigenvalue analysis has been done using different velocity profiles. It is shown that the effect of elasticity enhances the growth rate of shear flow driven Kelvin- Helmholtz (KH) instability.
The interplay between viscosity and elasticity not only enhances the growth rate but the spatial domain of the instability is also widened. The growth rate in various parameter space and the corresponding eigen functions are presented.
\end{abstract}
%\pacs{}
\maketitle
\section{Introduction}
In last few decades the importance of
dusty plasma   in space (e.g, in planetary rings, comet tails, interplanetary and interstellar clouds, in the vicinity of artificial satellites and space stations etc.,) and laboratory (technological plasma applications, fusion devices)  has increased making this area of research  interesting and useful.
The study of waves and instabilities in laboratory in presence of dust is  highly interesting because the additional charge species
is mutually connected with electron and ions via electromagnetic Lorentz forces\cite{pks}.
Since macroscopic dust particles can be visualized and tracked in particle level, dusty plasma is treated as a good experimental medium in laboratory to study phase transition\cite{morf}, transport properties and other collective phenomena\cite{piep,melz}. When macroscopic dust particles are added to an electron-ion plasma,
 large number of electrons are attached to a micron size dust surface  due to their higher mobility compared to ions. Thus dust act as a highly negative charged particle. It enables them to  strong electrostatic (Coulomb) interaction with the neighbouring dust particles so that the fluidity of dust particles becomes much less than that of normal electron-ion plasma. Hence, shear viscosity of dust fluid begins to play an important role  opposed to normal electron-ion plasma\cite{nose}. The strength of the Coulomb coupling is characterized by the coupling parameter $\Gamma = q_d^2/(k_B T_d a) $ where $q_d$ is the charge on the dust grains, $ a (\simeq n_d^{-1/3})$ is the average distance between them for density $n_d$, $T_d$ is the temperature of the dust component and $k_B$ is the Boltzmann constant\cite{ikej}. In the regime  $1 \leq \Gamma \leq \Gamma_c$ (a critical value beyond which system becomes crystalline) both viscosity and elasticity are equally important and therefore such plasma exhibits  visco-elastic behavior \cite{ikej,ichi}. When $\Gamma>\Gamma_{c}$, viscosity disappears and  only elasticity reigns over the system.   At high temperature, with the parameter  $\Gamma \ll 1$, the media exhibits  purely viscous
 effect but as the coupling parameter increases, Coulomb interaction between neighbouring particles becomes comparable to kinetic energy and hence the fluid also shows elastic property. Thus, strongly coupled plasmas cannot  be classified as purely elastic or purely viscous\cite{njp}.  Experimental observations  clearly demonstrated that a dusty plasma  with micron size negatively charged dust can readily go into a strongly coupled state and that the charged dust grains organize themselves into crystalline structures \cite{htho}. It is also shown that such a plasma  can support a transverse `shear mode'\cite{pkaw,pram}. At low temperature, potential energy easily overrules the kinetic energy of dust particles and hence dusty plasma fluid could posses memory dependent stress which leads to some elastic nature  along with its inherent viscous property. The KH instability is important in dusty plasma for understanding of various astrophysical phenomena where sheared dust flow naturally exists\cite{banr}. In a laboratory experiment application of external  dust shear flow may also be important to study the characteristics of KH instability.

In this paper, we have studied the linear stability analysis  of sheared dust flow  in presence of strong correlation between neighbouring dust particles. Unbounded parallel flow separated by a laminar shear layer could be unstable to small wavy disturbance depending on the nature velocity shear profile. This is a class of Kelvin Helmholtz instability that arises in parallel shear flows, where small-scale perturbations draw kinetic energy from the mean flow.  The effect of dust particles on the KH instabilities in electron-ion plasma for sheared ion flow  was studied before \cite{dngl}. In this work, the effect of both viscosity and elasticity  on the stability of sheared dust flow are studied with $\tanh$ type velocity profile. In recent past, it has been  reported in a strongly coupled yukawa liquid, strong coupling increases the growth rate of parallel shear flow instability with a model which does not have Galilean invariance\cite{ganu}. Here, the Generalized Hydrodynamic model is used with proper Galilean invariance [including the convective term $\left( {\bf v} \cdot \nabla\right)$ associated with $\tau_m$]  \cite{fren,sgnc} which provides a simple physical picture of the effects of strong correlations through the introduction of viscoelastic coefficients. This model is generally valid over a wide range of the coupling parameter.

\section{GH model and Stability Analysis}
  In this section we use standard fluid model of dusty plasma for studying low frequency ($\omega \ll k v_{th ~ e(i)}$) phenomena, where
  $\omega$ is the mode frequency, $k$ is the wave number, $ v_{th ~ e(i)}$ are electron and ion thermal velocity.
  Under this condition only massive dust dynamics is important.
  The dust fluid is considered here as homogeneous and incompressible so that density fluctuation can be ignored for simplicity. The Generalized Hydrodynamic equation for the dust fluid can be written as \cite{fren, pkaw}
        \bee
        \left \{1+ \tau_m \left (\fpar{}{t} +  {\bf v}\cdot \nabla\right) \right \} \left[\rho \left(\fpar{}{t}+{\bf v}\cdot \nabla\right){\bf v} +
        {q_d n_d}\nabla\varphi+ \nabla p \right ] = \fpar{\sigma_{ij}}{x_{j}}
        \label{deq}
        \ene
where  ${\bf v}$, $\rho(= M n_d)$, $n_d$, $p$, $\varphi$ are respectively  fluid velocity, mass density, dust mass, number density,  pressure and electrostatic potential. The parameter  $\tau_{m} = \eta / G$ is the relaxation time of the  medium with  viscosity coefficient $\eta$ and  rigidity modulus $G$. The strain tensor $\sigma_{ij}$ is given by
        \[
        \sigma_{ij} = \eta \left(\frac{\partial v_{i}}{\partial x_{j}} + \frac{\partial v_{j}}{\partial x_{i}}\right) +  \left(\xi- \frac{2}{3} \eta\right) \delta_{ij} \left( \nabla \cdot {\bf v}\right).
        \]
        where $\xi$ is the bulk viscosity coefficient.
 Earlier, GH model is being considered without the term
 $\tau_m \left( {\bf v} \cdot \nabla \right)$ \cite{ganu}. But, this is unfavorable for
 Galilean invariance in non-relativistic case. If one studies physics associated with equilibrium velocity shear this term have to be considered otherwise the analysis leads to  erroneous result.

 Let the equilibrium flow be along $x$-direction and it varies along $y$-direction so that ${\bf v}_0 = v_0(y) \hat e_{x}$. The total flow is the sum of equilibrium flow  and a small perturbation:
 \[
 {\bf v}(x,y,t) = [v_0(y) + v_{x}(x,y,t)] \hat e_{x}+ v_{y}(x,y,t) \hat e_{y}.
 \]
 Linearizing Eq. (\ref{deq}) around equilibrium flow $v_0$, scalar component equations can be written as
        \bee
        \left \{1+ \tau_m \left (\fpar{}{t} +   v_0 \fpar{}{x}\right) \right \} \left[ \left(\fpar{}{t}
        + v_0 \fpar{}{x}\right)v_{x} + v_y \nder{}{v_0}{y}+ \frac{1}{\rho} \fpar{p}{x}+\frac{q}{M} \fpar{\varphi}{x}\right ] = \nu \nabla^2 v_{x} + \left(\frac{\xi}{\rho} + \frac{\nu}{3}\right) \frac{\partial}{\partial x}\left( \nabla \cdot {\bf v}\right),
        \label{cex}
        \ene

        \bee
        \left \{1+ \tau_m \left (\fpar{}{t} +   v_0 \fpar{}{x}\right) \right \} \left[ \left(\fpar{}{t}
        + v_0 \fpar{}{x}\right)v_y + \frac{1}{\rho} \fpar{p}{y} + \frac{q}{M} \fpar{\varphi}{y}\right] = \nu \nabla^2 v_{y} + \left(\frac{\xi}{\rho} + \frac{\nu}{3}\right) \frac{\partial}{\partial y}\left( \nabla \cdot {\bf v}\right).
        \label{cey}
        \ene

Differentiating equation(\ref{cex}) with respect to y, and equation(\ref{cey}) with respect to x, we get
        \beea
        \left \{1+ \tau_m \left (\fpar{}{t} +   v_0 \fpar{}{x}\right) \right \} \left[ \left(\fpar{}{t}
        + v_0 \fpar{}{x}\right) \frac{\partial v_{x}}{\partial y} + v_0^{''}v_y + \frac{1}{\rho} \frac{\partial^2 p}{\partial y \partial x}+\frac{q}{M} \frac{\partial^2 \varphi}{\partial y \partial x}\right ]\nonumber\\ + \tau_m v_{0}^{'} \fpar{}{x} \left[ \left( \fpar{v_x}{t} + v_0 \fpar{v_x}{x}\right) + v_{0}^{'} v_{y} + \frac{1}{\rho}\fpar{p}{x} + \frac{q}{M} \fpar{\varphi}{x} \right] = \nu \nabla^2 \frac{\partial v_{x}}{\partial y} + \left(\frac{\xi}{\rho} + \frac{\nu}{3}\right) \frac{\partial^2}{\partial y \partial x}\left( \nabla \cdot {\bf v}\right),
        \label{dex}
        \enea

\bee
        \left \{1+ \tau_m \left (\fpar{}{t} +   v_0 \fpar{}{x}\right) \right \} \left[ \left(\fpar{}{t}
        + v_0 \fpar{}{x}\right) \frac{\partial v_y}{\partial x} + \frac{1}{\rho} \frac{\partial^2 p}{\partial x \partial y}+\frac{q}{M} \frac{\partial^2 \varphi}{\partial x \partial y}\right] = \nu \nabla^2 \frac{\partial v_{y}}{\partial x} + \left(\frac{\xi}{\rho} + \frac{\nu}{3}\right) \frac{\partial^2}{\partial x \partial y}\left( \nabla \cdot {\bf v}\right)
        \label{dey}
        \ene

 where kinematic viscosity coefficient $\nu = \eta/\rho$.
Then subtracting equation(\ref{dey})from the equation(\ref{dex})  we obtain
\beea
\left \{ 1 + \tau_m \left( \fpar{}{t} +   v_0 \fpar{}{x}\right) \right \} \left[ \left(\fpar{}{t} + v_0 \fpar{}{x}\right) \left( \fpar{v_x}{y} - \fpar{v_y}{x} \right) + v_{0}^{''} v_{y} \right]\nonumber\\ + \tau_m v_{0}^{'} \fpar{}{x} \left[ \left( \fpar{}{t} + v_0 \fpar{}{x}\right) v_x + v_{0}^{'} v_{y} + \frac{1}{\rho}
\fpar{p}{x} - \frac{Ze}{M} \fpar{\varphi}{x} \right] = \nu \nabla^2 \left( \fpar{v_x}{y} - \fpar{v_y}{x}\right)
\label{veq}
\enea
where  dust charge $q = -Ze$, $Z$ is the number of electrons on each dust and $v_0^{'}$ and $v_0^{''}$ are respectively $\nder{}{v_0}{y}$ and $\nder{2}{v_0}{y}$.
%%%%%%%%%%

%\beea
%\left \{ 1 + \tau_m \left( \fpar{}{t} +   v_0 \fpar{}{x}\right) \right \} \left[ \left(\fpar{}{t} + v_0 \fpar{}{x}\right) \nabla^2 \psi - v_{0}^{''} \fpar{\psi}{x} \right]\nonumber\\ + \tau_m v_{0}^{'} \fpar{}{x}
%\left[ \left( \fpar{}{t} + v_0 \fpar{}{x}\right) \fpar{\psi}{y} - v_{0}^{'} \fpar{\psi}{x} + \frac{1}{\rho}
%\fpar{p}{x} + \frac{Ze}{M} \fpar{\varphi}{x} \right] = \nu \nabla^4 \psi
%\label{veq}
%\enae
 %As shear flow dynamics of dust does not introduce compressibility in the system and hence density fluctuation, we are considering incompressible dusty plasma in our study. Also, we are considering cold dust plasma i.e, random thermal motions of dust are ignored. Kelvin-Helmholtz instability is an electrostatic instability, so our case does not take care electromagnetic situation. In electrostatic media, electric field fluctuation originates owing to the density fluctuation (compressible phenomena) of charge particles which is expressed in poission's equation.

 To study the dynamics of pure shear flow in a dusty plasma we are neglecting the effects of density fluctuation by considering an incompressible medium. Also, we are assuming cold dust particles i.e, random thermal motion of dust is ignored. In this article, only electrostatic KH instability will be studied. In electrostatic media, electric field fluctuation originates owing to the density fluctuation (compressible phenomena) of charge particles which are mathematically connected through Poisson's equation. Hence, the pressure and the electric field perturbation terms in equation(\ref{veq}) will not contribute further.
 The incompressibility condition is given by
 \bee
 \fpar{v_x}{x} + \fpar{v_y}{y} = 0.
 \label{com}
 \ene
A possible solution of Eq. (\ref{com}) in terms of a stream function $\psi$ may be written as $v_x = \fpar{\psi}{y},  v_y = -\fpar{\psi}{x}$.
Hence the equation(\ref{veq}) can be written as,
\beea
\left \{ 1 + \tau_m \left( \fpar{}{t} +   v_0 \fpar{}{x}\right) \right \} \left[ \left(\fpar{}{t} + v_0 \fpar{}{x}\right) \nabla^2 \psi - v_{0}^{''} \fpar{\psi}{x} \right] + \tau_m v_{0}^{'} \fpar{}{x} \left[ \left( \fpar{}{t} + v_0 \fpar{}{x}\right) \fpar{\psi}{y} - v_{0}^{'} \fpar{\psi}{x} \right] = \nu \nabla^4 \psi.
\label{veq}
\enea

The problem considered here is linear and inhomogeneous in $y$ so  any arbitrary disturbance may be decomposed into normal
modes as
\[
\psi(x,y,t) = \phi(y) e^{i(k x -\omega t)},
\]
where $\omega = k c$, and $c$ is the phase velocity of the wave.  Using this normal mode form, equation(\ref{veq}) can be written in a dimensionless form as,
\beea
(D^2-k^2)^2 \phi(y) = i k R \left[ \left \{  1 + i k \tau_m \left( v_0 - c \right) \right \} \left
\{ \left( v_0 -c \right) \left( D^2 - k^2 \right)  - v_0^{''} \right \} + i k \tau_m v_{0}^{'} \left \{ \left( v_0 - c \right) D - v_{0}^{'} \right \} \right] \phi(y)
\label{GHOS}
\enea
where, $D$ denotes ${d}/{dy}$, Reynolds number $R= v_0 L \rho/\eta$ and $L$ is equilibrium shear length scale.
This equation describes the visco-elastic stability of normal modes of parallel shear flow in strongly coupled dusty plasma. In the limit $\tau_m = 0$, this equation leads to the  celebrated  Orr-Sommerfeld equation \cite{pant,dzre} which examines the behavior of small disturbances in the parallel flow of an incompressible viscous fluid.

Now we consider the following discontinuous steady velocity profile which help us to treat the problem analytically
\[
v_0(y) = y/|y| ; ~~~~~~~~~~~ -\infty \leq y \leq \infty
\]
The above velocity profile shows that at $y = 0$,   $v_0$ has a sudden jump but, in the regions $y>0$ and $y<0$, the profile is continuous and constant. Only at $y = 0$, both first and second derivatives of velocity exist. In weakly coupled limit,
 the stability of this type of piecewise continuous velocity profile in a viscous incompressible fluid was analytically studied and an
 instability \cite{Draz} was predicted.
For the regions $y<0$ and $y>0$, the Generalized Hydrodynamic Orr-Sommerfeld equation (\ref{GHOS}) reduced to
\bee
(D^2-k^2)^2 \phi = i k R \left \{  1 + i k \tau_m \left( \mp 1 - c \right) \right \} \left( \mp 1 -c \right) \left( D^2 - k^2 \right) \phi
\label{oslm}
\ene
Note here that  $v_{0}^{'}$ and $v_{0}^{''}$ do not appear in the above equation. But, the effect
of sudden jump in the velocity profile at $y = 0$ would appear through the boundary condition. The most general solution of equation (\ref{oslm}) satisfying the boundary condition at infinity is of the form
\beea
\phi =  A e^{-ky} + B e^{-\beta_1 y} ~~~~ \left( y>0 \right)\nonumber\\
               C e^{ky} + D e^{\beta_2 y}  ~~~~ \left( y<0 \right)
\label{sol}
\enea

where
\[
\beta_1 = \left[ k^2 - ikR \left( c - 1 \right) \left\{ 1 - i \tau_m k \left( c - 1 \right)\right\} \right]^{1/2},\;\;
\beta_2 = \left[ k^2 - ikR \left( c + 1 \right) \left\{1 - i \tau_m k \left( c + 1 \right)\right\} \right]^{1/2}.
\]
On integrating  equation (\ref{GHOS}) successively across the discontinuity of velocity profile for infinitesimal regions, we get the boundary conditions
\beea
\left[ \phi \right] = 0, \\
\left[ D \phi\right]=0, \\
\left[ \left(D^2 + \beta^2\right) \phi + \frac{R
 \tau_m k^2}{2} \left( v_0 - c\right)^2 \phi \right] = 0,\\
\left[ \left(D^2 - \beta^2\right) D \phi \right] =0,
\label{bc}
\enea
where
\[
\beta = \left[ k^2 - ikR \left( c - U_0 \right) \left( 1 - i \tau_m k \left( c - v_0 \right)\right) \right]^{1/2}.
\]
The boundary conditions at $y = 0$ gives four homogeneous linear equations in $A,B, C$ and  $D$  as indicated in Eq. (\ref{sol}).
A non zero solution for these set of equations exists if and only if their discriminant is zero.
 A straightforward algebra result  to the eigenvalue condition
\[
k^2+\beta_1^2 + \beta_2^2 - \beta_1 \beta_2 + k(\beta_1 + \beta_2) = \Delta\left( \frac{\beta_1 - \beta_2}{\beta_1 + \beta_2}\right),
\]
where $\Delta= R \tau_m c k^2$.

                \begin{figure}
                    \centering
                    \includegraphics[width=3.0in,height=2.5in]{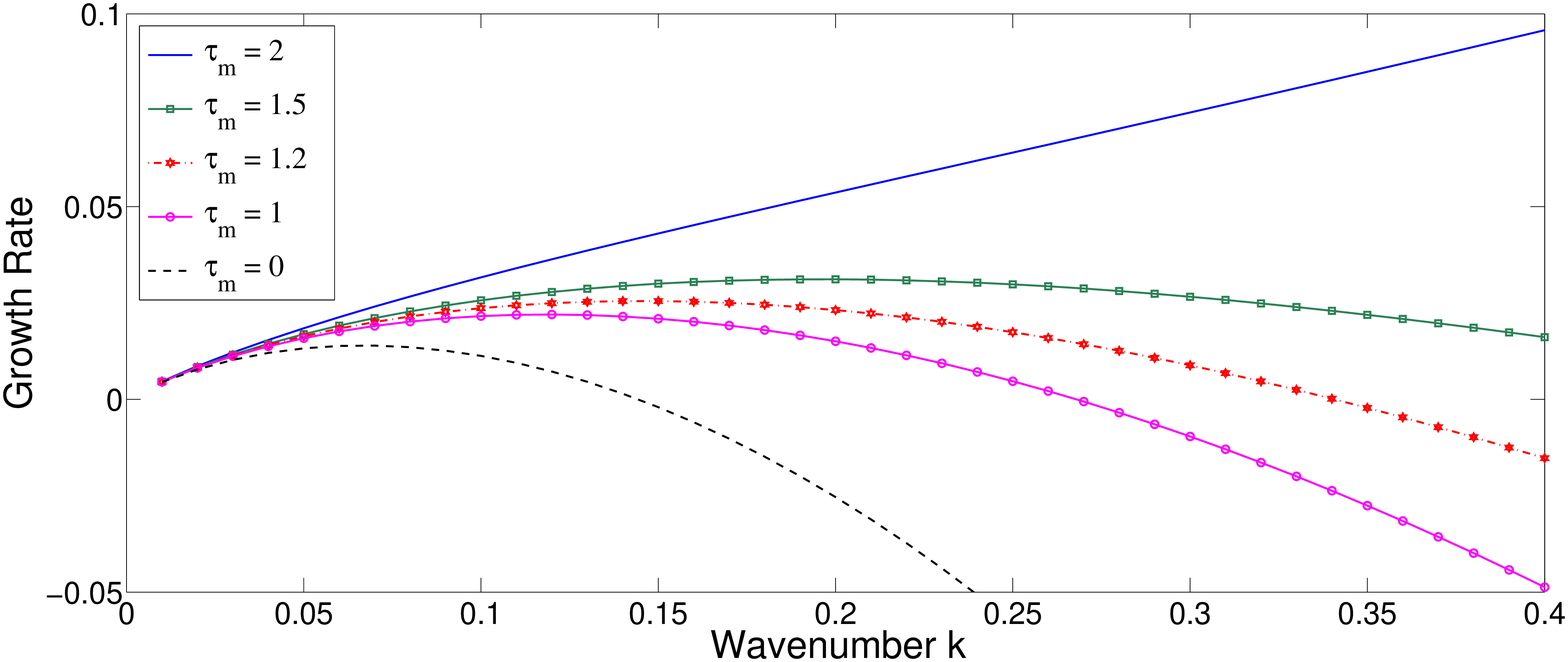}
                    \caption{(Color online)Variation of growth rate with wave number is plotted for step shear profile with Reynolds Number $R = 1$ and for different values of $\tau_m$ enlisted in legend. With increase of relaxation time $\tau_m$, growth rate also increases for each k.}
                    \label{anl}
                \end{figure}
Figure (\ref{anl}) shows a plot of growth rate vs. wave number for various values of $\tau_m$. For $\tau_m = 0$, the dotted curve shows the result in a weakly coupled limit.  This figure clearly indicates that increase of relaxation time enhances instability. Strong coupling between dust particles ($\Gamma$) increases relaxation time ($\tau_m$) \cite{fgl} and thus enhances the growth rate of KH instability.
\newpage
\section{Eigenvalue Analysis}
In the previous section, non-local analysis has shown that strong coupling between neighboring dust particles enhances the instability of parallel sheared dust flow. As a step profile is not a realistic profile,  a $tanh$ type profile is considered which is widely treated in both experimental and simulation studies. The expression of  such a profile is given by
\bee
v_0 = \bar{v}_{0} \tanh(y/L),
\nonumber
\ene
where $L$ is typically the velocity shear inhomogeneity length and $\bar{v}_{0}$ is the magnitude of velocity far away from shear region. Hence the velocity smoothly varies from $-\bar{v}_{0}$ to the value $\bar{v}_{0}$ in the width of shear region $2L$. We have done matrix eigenvalue analysis  of the differential equation (\ref{GHOS}) for the above mentioned profile using standard eigenvalue subroutine ({\bf eig})in MATLAB after properly discretization of the said equation with standard finite difference discretization scheme.  Following central difference scheme is used for the purpose of discretization.

\begin{center}
\beea
\nder{4}{\phi}{y} = \frac{\phi_{i+2}-4\phi_{i+1}+6\phi_i-4\phi_{i-1}+\phi_{i-2}}{\Delta^4}\nonumber\\
\nder{2}{\phi}{y} = \frac{\phi_{i+1}-2\phi_{i}+\phi_{i-1}}{\Delta^2}~~~~~~~~~~~~~~~~~~~\nonumber\\
\nder{}{\phi}{y}  = \frac{\phi_{i+1}-\phi_{i-1}}{2\Delta}~~~~~~~~~~~~~~~~~~~~~~~~~~\nonumber
\enea
\end{center}
where $\Delta$ is the grid spacing.
After a few algebraic steps, the linearized fourth order equation(\ref{GHOS}) reduces to polynomial eigenvalue problem in $\omega$as

\bee
\left[ A_0 - \omega A_1 - \omega^2 A_2 \right] \phi = 0,
\ene
where $A_i$'s are the matrix elements in above equation and $\omega = kc $ ($c$ is the phase velocity). The polynomial eigenvalue problem can be changed into general eigenvalue problem using the dummy variable
$\chi=\omega \phi$. Hence, the new eigenvalue problem is
\bee
      \left
       (\begin{array}{lr}
       A_0 & Z\\ Z & I\\
         \end{array}\right)
         \left(\begin{array}{c}\phi \\ \chi\\
         \end{array}\right) = \omega
         \left
       (\begin{array}{lr}
       A_1 & A_2\\ I & Z\\
         \end{array}\right)
         \left(\begin{array}{c}\phi \\ \chi\\
         \end{array}\right),
         \label{mt}
\ene
 where $I$ is identity matrix and $Z$ is a null matrix. This trick simplifies original polynomial eigenvalue problem
 into a simple and well known matrix eigenvalue problem as
 \[
 R \bar{\phi} = \omega S \bar{\phi}
 \]
 where
 \bee
 R = \left(\begin{array}{lr}A_0 & Z\\ Z & I\\\end{array}\right);~~\nonumber\\
 S = \left(\begin{array}{lr}A_1 & A_2\\ I & Z\\\end{array}\right);~~\nonumber\\
 \bar{\phi} = \left(\begin{array}{c}\phi \\ \chi\\\end{array}\right).\nonumber
 \ene
 Now, we can use eig subroutine to solve the eigenvalue equation.
 We have calculated the imaginary part of eigenvalues $\omega$, the
positive value of which  indicates the growth rate of the KH mode. First we should validate our code with respect to existing results. In weakly coupled limit($\tau_m \ll 1$), equation(\ref{GHOS}) reduces to the well known Orr-Somerfeld equation which has been thoroughly studied in the last century.
\begin{figure}
       \centering
       \includegraphics[width=4in,height=3in]{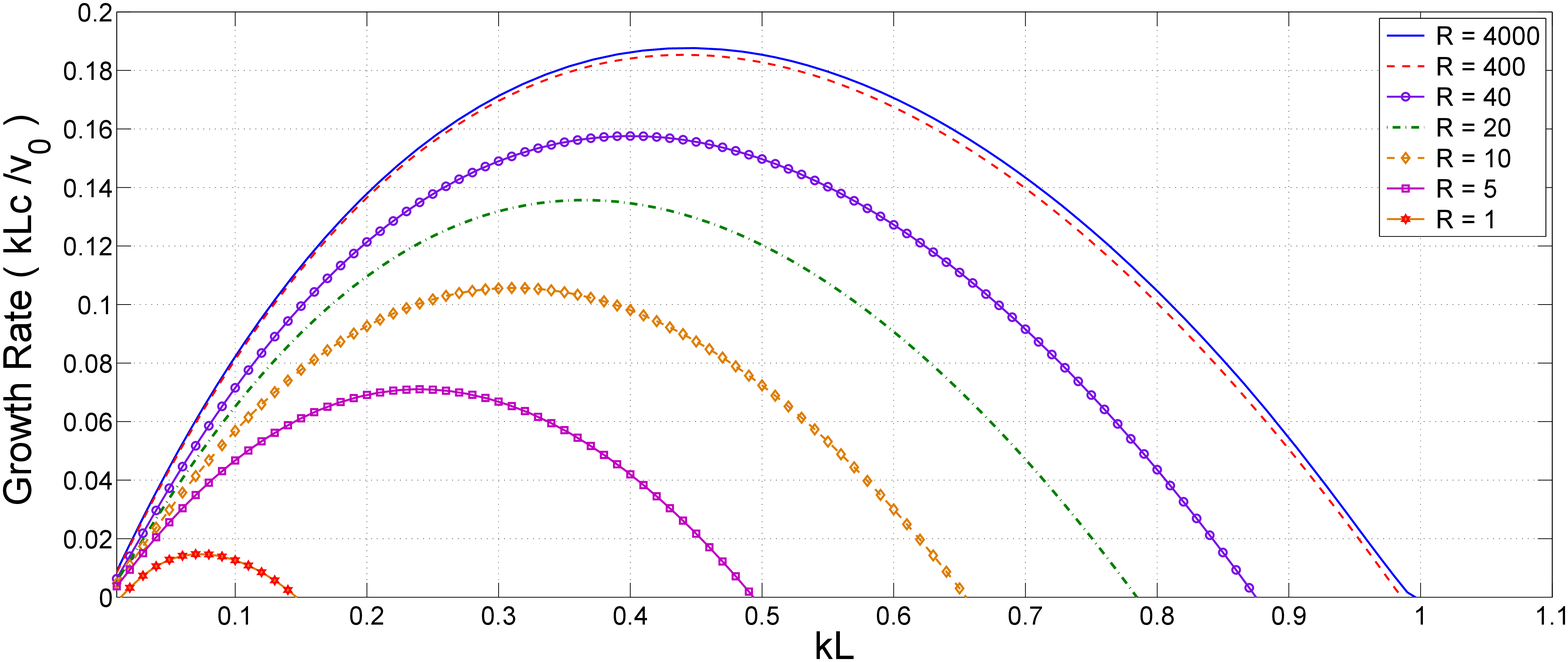}
       \caption{(Color online)Growth rate for different Reynolds Number(R) is plotted against wavenumber  in weakly coupled limit ($\tau_m = 0$). Growth rate($kc$) is calculated in unit of $v_0/L$ and wavenumber($k$) in unit of $1/L$. Different colors indicate different values of R. Viscous stabilization is clearly seen for small R and its large value proceeds towards inviscid limit $R\rightarrow \infty$.}
       \label{vald}
\end{figure}
In  figure (2), we have plotted the growth rate against wave number for different values of Reynolds number $R$. These results agree
with the results of Fig. 1. in  Ref. \cite{betc}.
 The code also shows that instability of $tanh$ velocity profile increases as viscosity decreases and for very large value
 of Reynolds number $R$, the result resemble to those obtained in the inviscid limit.

\begin{figure}
  \begin{center}
    \begin{tabular}{cc}
      \resizebox{90mm}{!}{\includegraphics{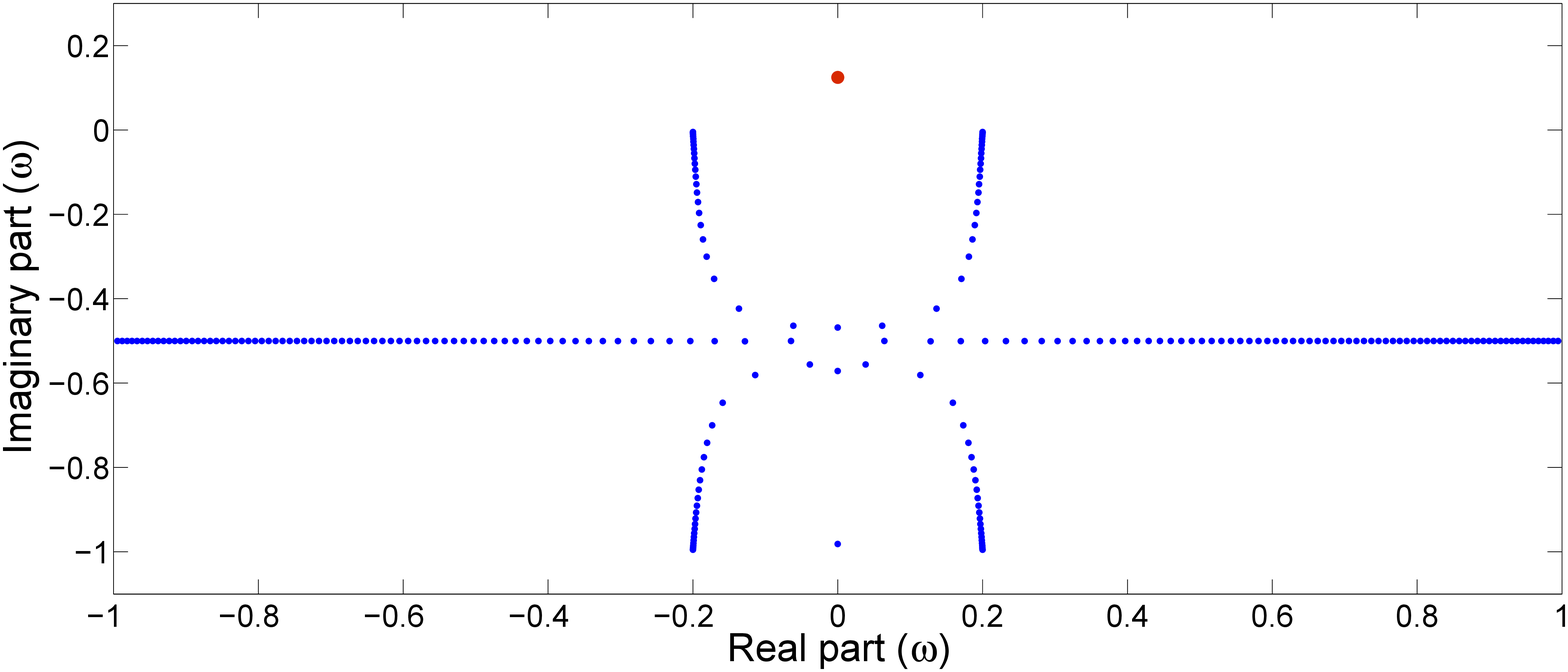}} &
      \resizebox{90mm}{!}{\includegraphics{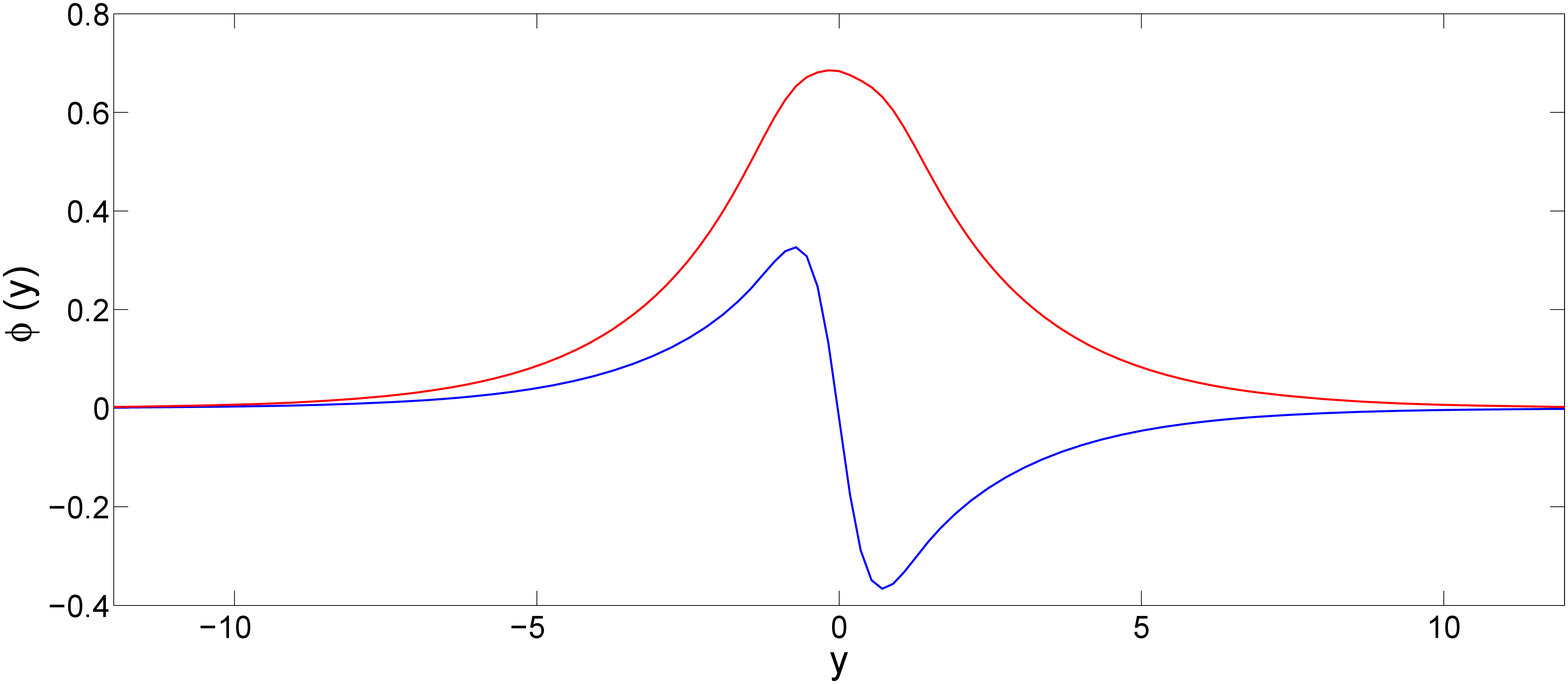}} \\
      \end{tabular}
    \caption{(Color online)Eigenvalues are shown in complex plane for $\tau_m = 1$ and $R = 10$ in the left graph. Red(big) dot represents the only unstable mode which is purely imaginary. Real(blue(lower) line) and imaginary(red(upper) line) parts of eigenfunction corresponding to unstable mode are plotted in the right graph.}
   \label{test4}
  \end{center}
\end{figure}

 After this bench marking result, we investigate the growth rate of  KH instability for different values of viscosity and relaxation time.
The fact is that unstable mode has no real part i.e., it lies on the imaginary axis in the complex plane. In  figure (\ref{test4}), eigenvalues in the complex plane have been plotted for $R =1,10 $ and $\tau_m = 1$ and the corresponding localized eigenfunctions are also shown. In figures (\ref{gw1})-(\ref{gw3}), we  have  also shown growth rate vs. wave number curve for different values of $\tau_m$. As Reynolds number increases, growth rate for different $\tau_m$ values also increases.

\begin{figure}
  \begin{center}
    \begin{tabular}{cc}
      \resizebox{90mm}{!}{\includegraphics{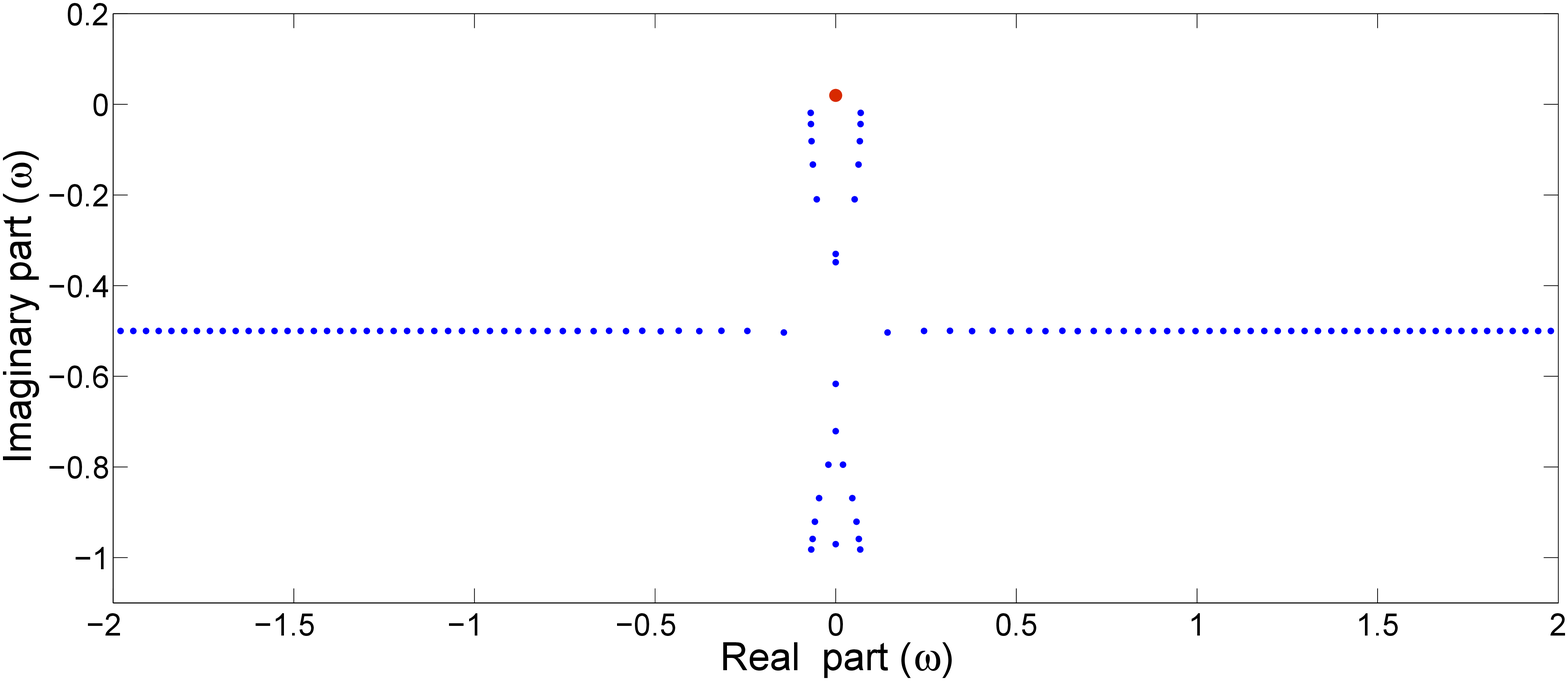}} &
      \resizebox{90mm}{!}{\includegraphics{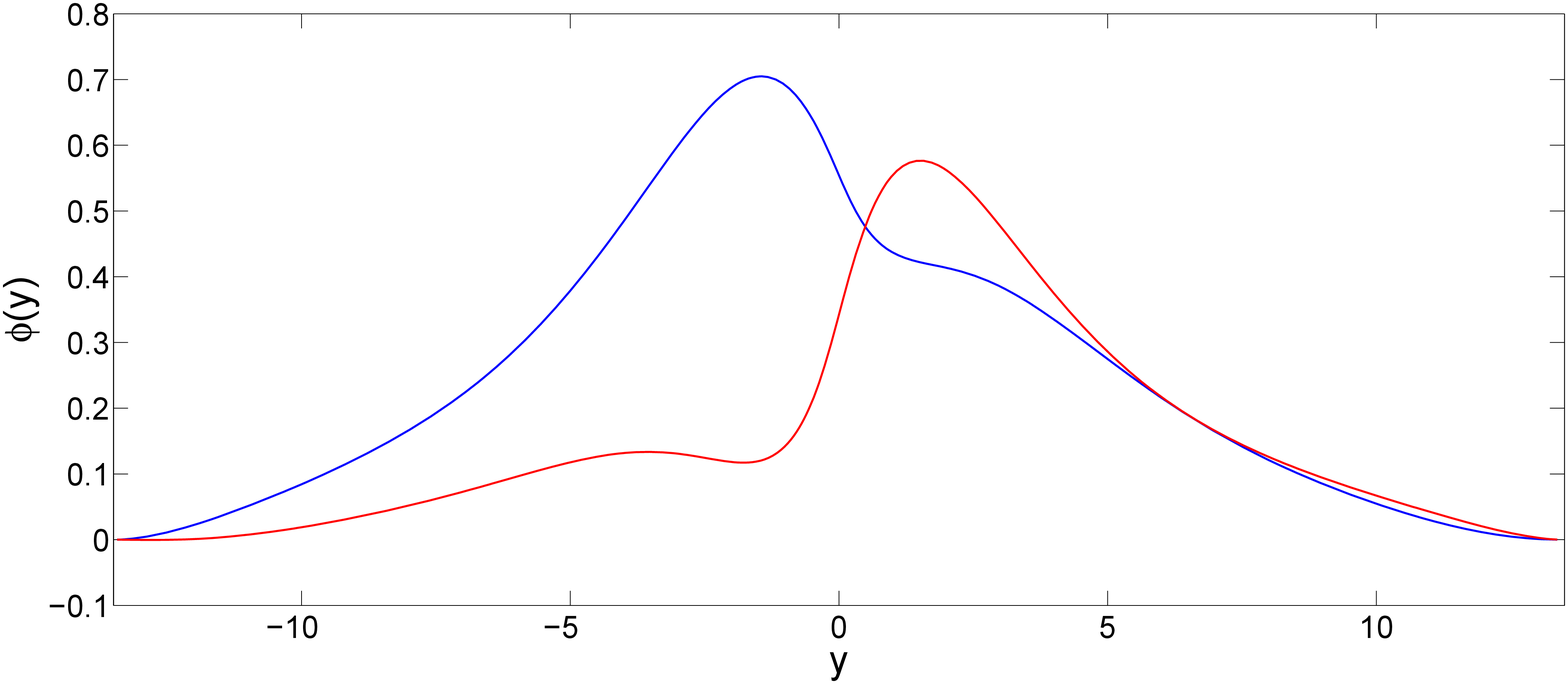}} \\
      \end{tabular}
    \caption{(Color online)Eigenvalues are shown in complex plane for $\tau_m = 1$ and $R = 1$ in the left graph. Red(big) dot represents the only unstable mode which is purely imaginary. Real(blue(upper) line) and imaginary(red(lower) line) parts of eigenfunction corresponding to unstable mode are shown in the right graph.}
   \label{test5}
  \end{center}
\end{figure}

\begin{table*}[h]
\begin{center}{\footnotesize
\begin{tabular}{| c | c | c | l |}
\hline
~~~Reynolds number(R) ~~~ & ~~~~~~~~$\tau_m$ ~~~~~~~&~~~~~~Max Growth rate ~~~~~~~  &~~~~~~~ Max Growth rate  \\
                      &                           & with $\tau_m ({\bf v}\cdot \nabla) $   &~~~~~~~ without $\tau_m ({\bf v}\cdot \nabla)$ \\
\hline
 $  $  &    $ 0.0 $ &   $ 0.01474 $    &  $~~~~~~~~~ 0.01474 $ \\
 $  $  &    $ 1.0 $ &   $ 0.02256 $    &    $~~~~~~~~~  0.01499 $ \\
 $  $  &    $ 2.0 $ &   $ 0.07569 $   &  $~~~~~~~~~  0.01520 $ \\
 $ 1 $  &    $ 3.0 $ &   $ 0.1069 $    &    $~~~~~~~~~  0.01575 $ \\
 $  $  &    $ 5.0 $ &   $ 0.2267 $    &    $~~~~~~~~~  0.01621 $ \\
 $  $  &    $ 8.0 $ &   $ 0.3197 $    &    $ ~~~~~~~~~ 0.01690 $ \\
 $  $  &    $ 10.0 $ &   $ 0.3545 $    &    $~~~~~~~~~  0.0179$ \\
\hline
\end{tabular}}
\end{center}
\caption{ Comparison of growth rates for Galilean invariant and non-invariant GH model}
\label{table:cmpre}
\end{table*}
 The Generalized Hydrodynamic model is becoming  an inevitable tool to study the effect of strong coupling between dust particles on different waves and instabilities in a dusty plasma. In many cases, proper  model was not taken into consideration.
For the study of Kelvin-Helmholtz instability where equilibrium shear flow
plays an important role, it is necessary to consider a proper
Galilean invariant GH model. With the convective terms associated with relaxation
time taken into account
the growth rate of unstable mode is plotted against wave number in fig. (\ref{gw1}) for $ R = 1 $ in both cases of including or excluding the term $\tau_m \left( {\bf v} \cdot \nabla \right)$.
These two figures clearly indicate that the proper Galilean invariant form of the GH model makes a drastic change in growth rate using $tanh$ type velocity profiles. A comparison of growth rates is given in tabular form for different
 $\tau_m$ values. It is also observed that the limiting value of k beyond which instability vanishes  also changes for different values of relaxation time $\tau_m$.

\begin{figure}
   \begin{center}
   \begin{tabular}{lr}
     \resizebox{90mm}{!}{\includegraphics[width=4in,height=3.0in]{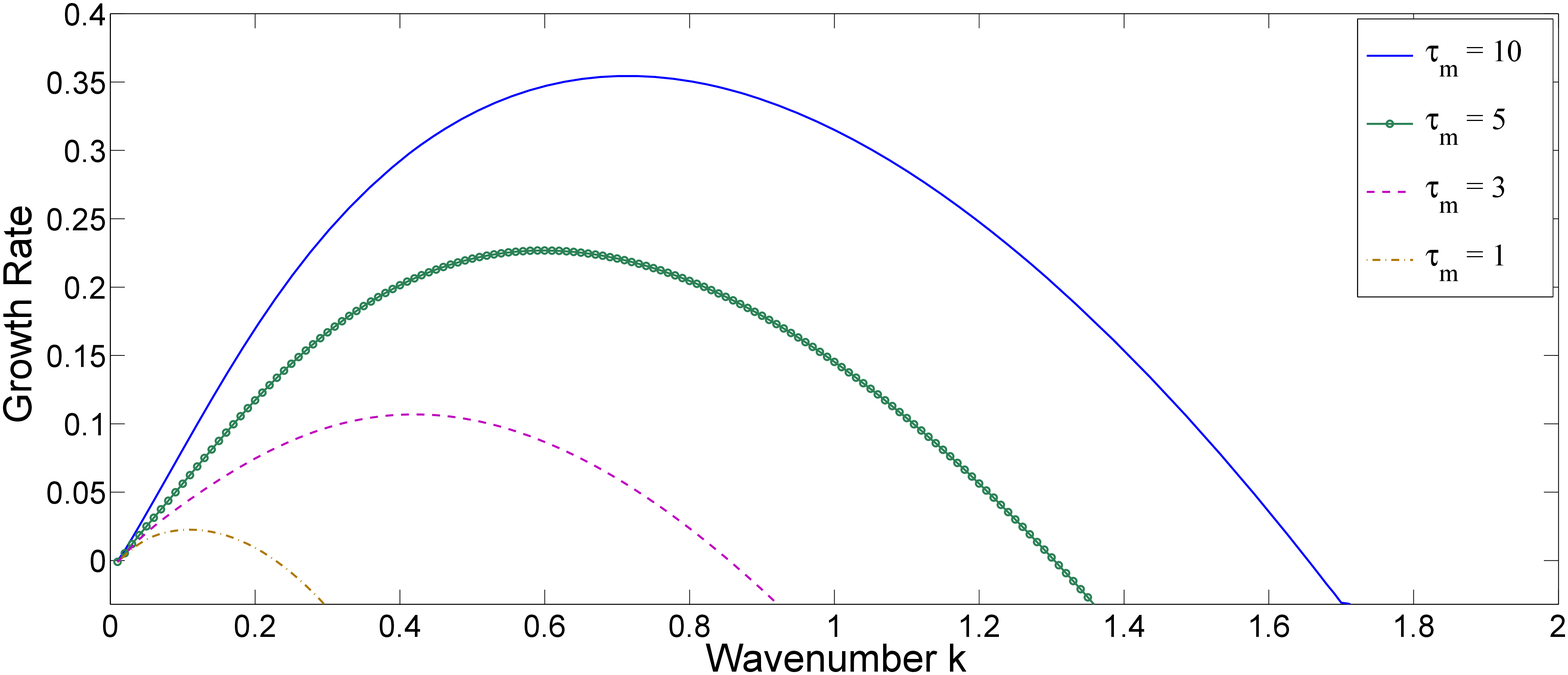}} &
     \resizebox{90mm}{!}{\includegraphics[width=4in,height=3.0in]{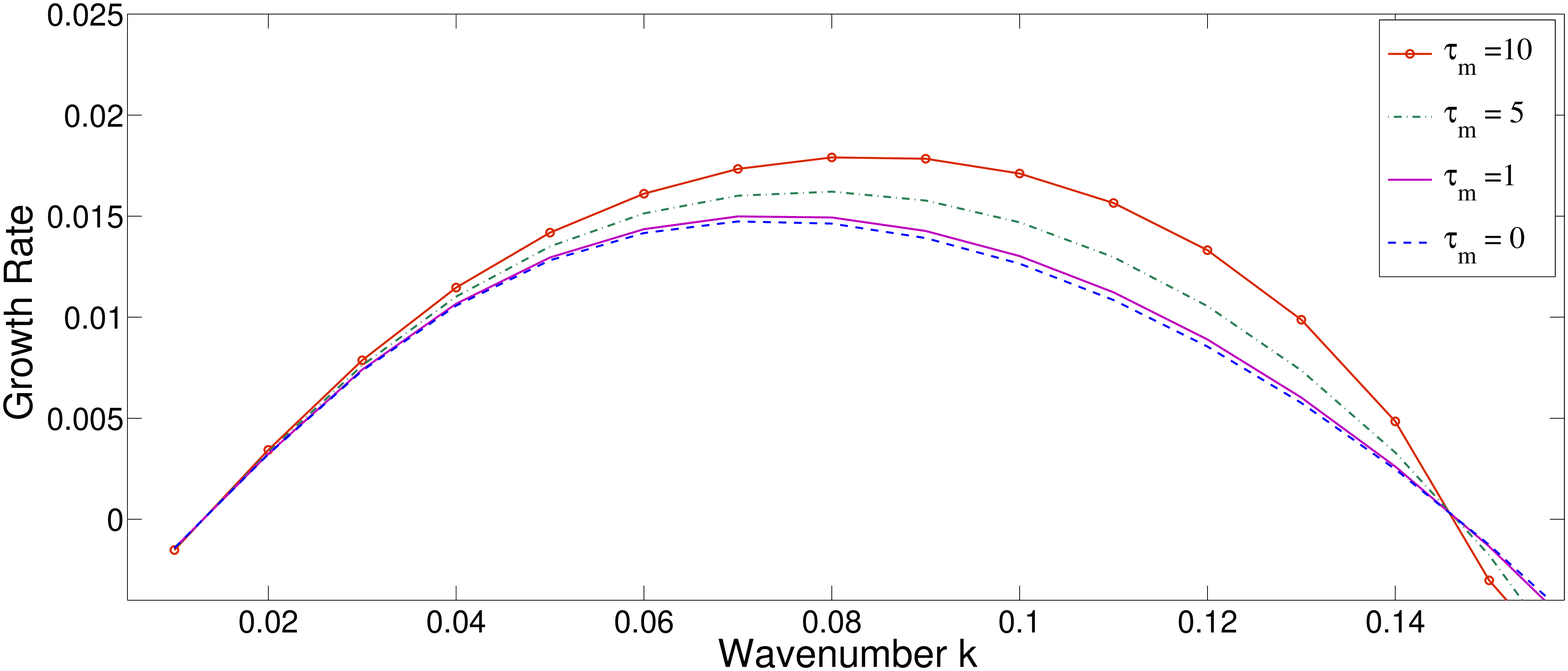}}
     \end{tabular}
     \caption{ (Color online)Growth rate vs. wavenumber curves are shown for different values $\tau_m$ for two different cases -- left one with taking $\tau_m \left( {\bf v} \cdot \nabla \right)$ term in GH model and right one without that term. $R = 1$.}
     \label{gw1}
   \end{center}
 \end{figure}
\begin{figure}
   \begin{center}
    \begin{tabular}{ll}
     \resizebox{90mm}{!}{\includegraphics[width=4in,height=3.0in]{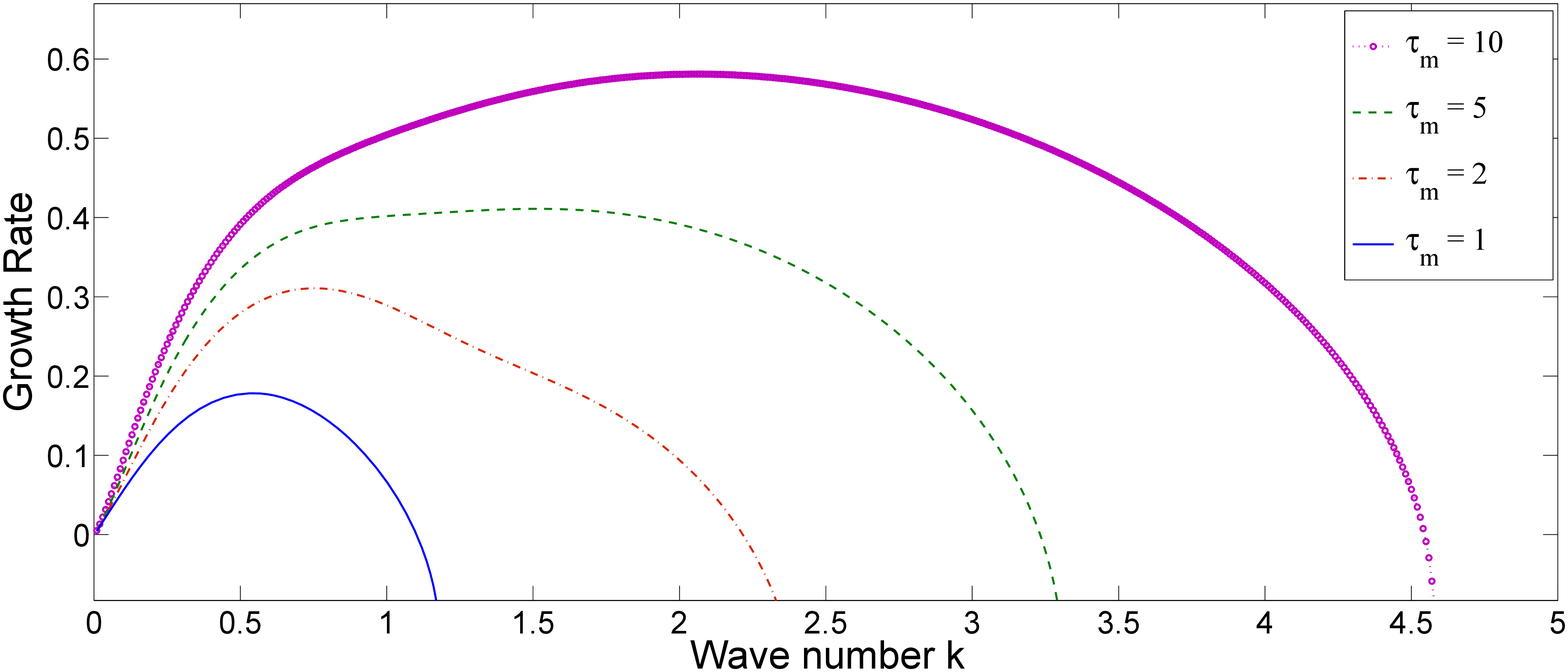}} &
     \resizebox{90mm}{!}{\includegraphics[width=4in,height=3.0in]{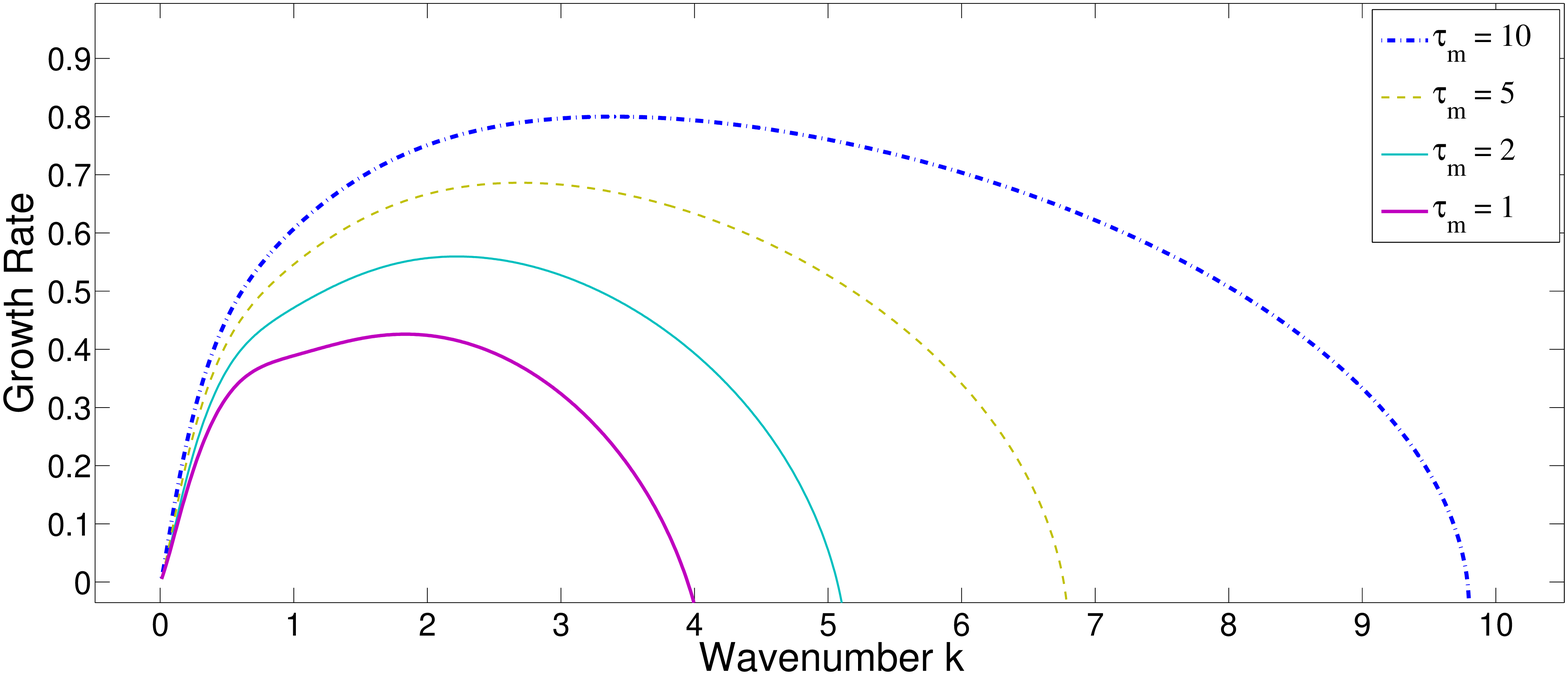}}
     \end{tabular}
     \caption{(Color online)For $R = 5$ (left panel) and $R = 10$ (right panel) and different values of $\tau_m$, variation of Growth rate with k is shown. }
     \label{gw2}
   \end{center}
 \end{figure}
\begin{figure}
   \begin{center}
    \begin{tabular}{rr}
     \resizebox{90mm}{!}{\includegraphics[width=4in,height=3.0in]{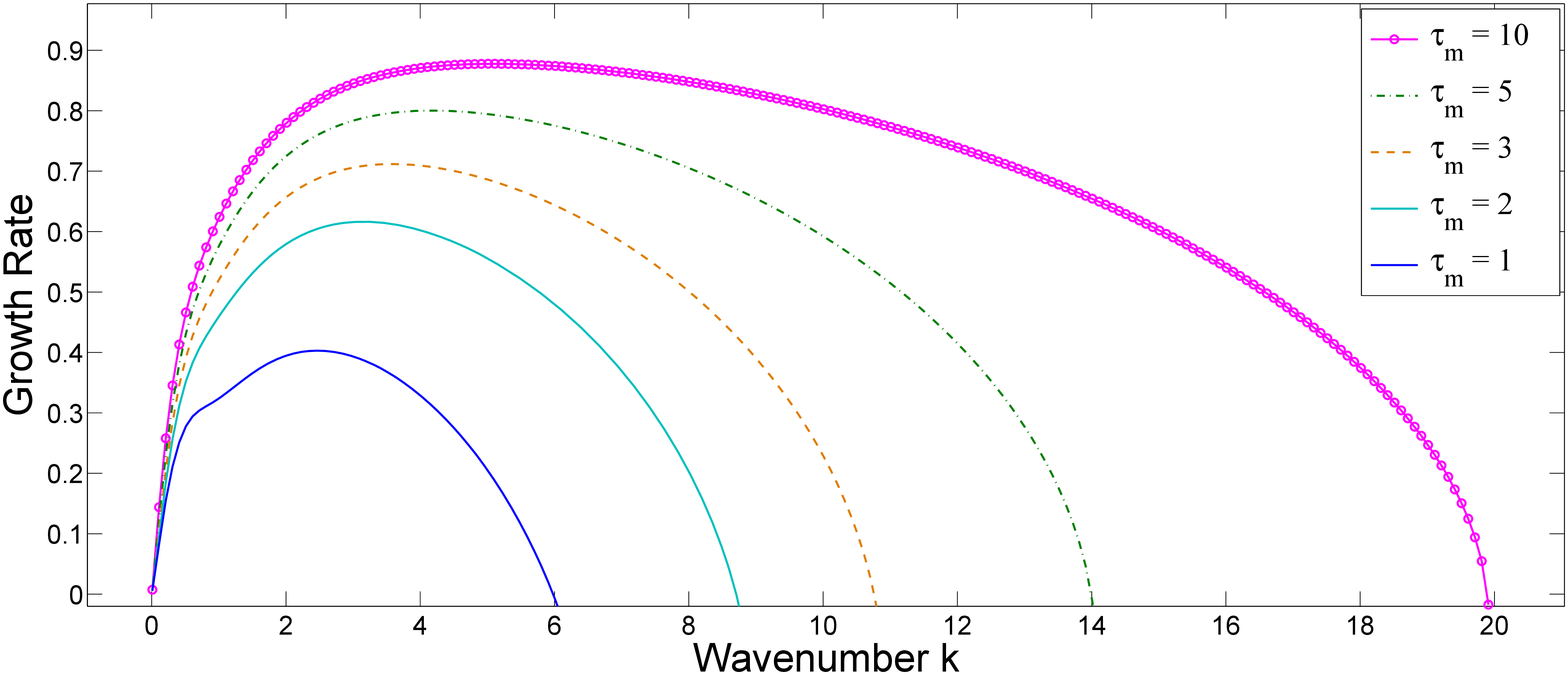}} &
     \resizebox{90mm}{!}{\includegraphics[width=4in,height=3.0in]{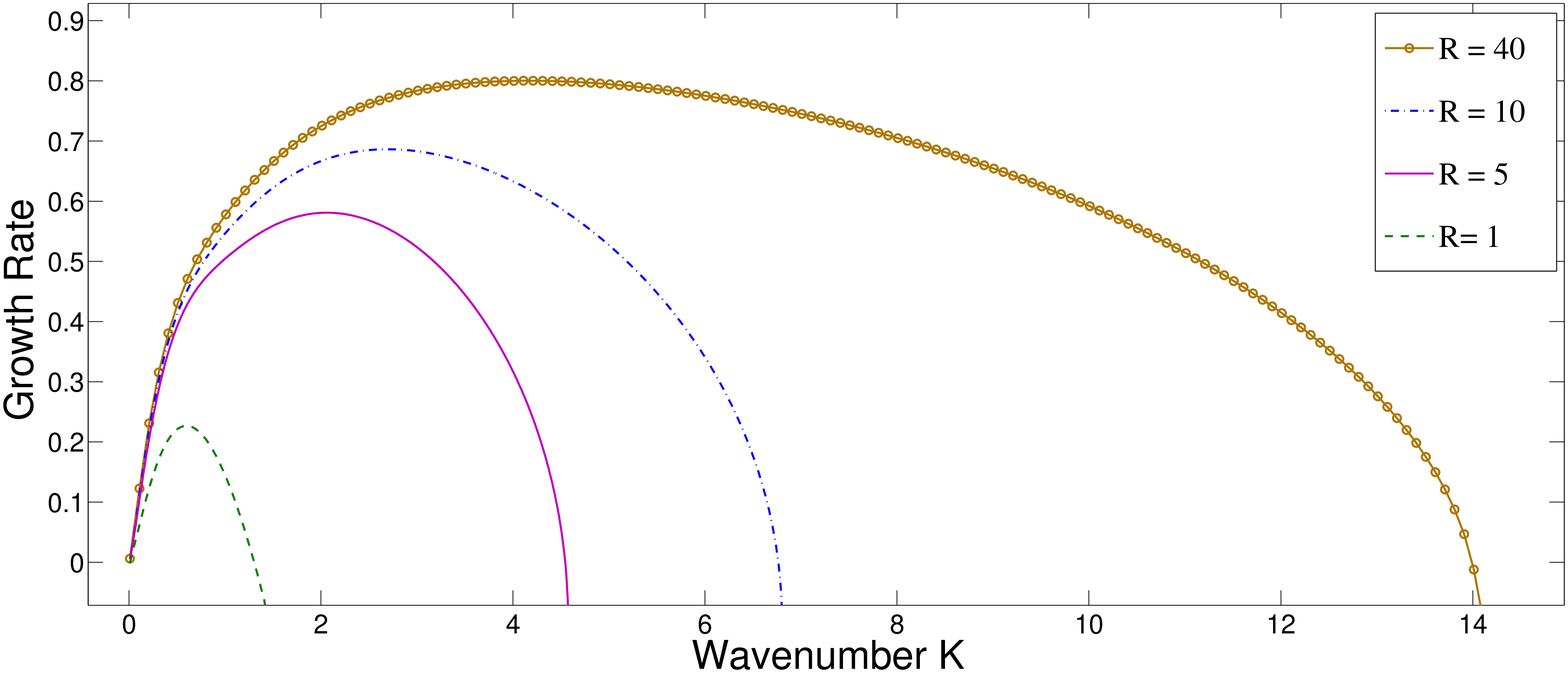}}
     \end{tabular}
     \caption{(Color online)In left figure, growth rate vs. $k$ is plotted for $R = 40$ and different $\tau_m$. Right figure shows growth rate variation with $k$ for  $\tau_m = 5$ but different Reynolds numbers.}
     \label{gw3}
   \end{center}
 \end{figure}
%%
%\begin{figure}
%\centering
%\includegraphics[width=4in,height=3.0in]{cont1-fig7.eps}
%\caption{Contour plot in 2D plane of $k$ and $R$ for  weakly coupled limit $\tau_m = 0$ and  for  relaxation time $\tau_m = 1$. Blue lines show former case and magenta lines the latter one.}
%\label{sp1}
%\end{figure}
\begin{figure}
   \begin{center}
    \begin{tabular}{rr}
     \resizebox{90mm}{!}{\includegraphics[width=4in,height=3.0in]{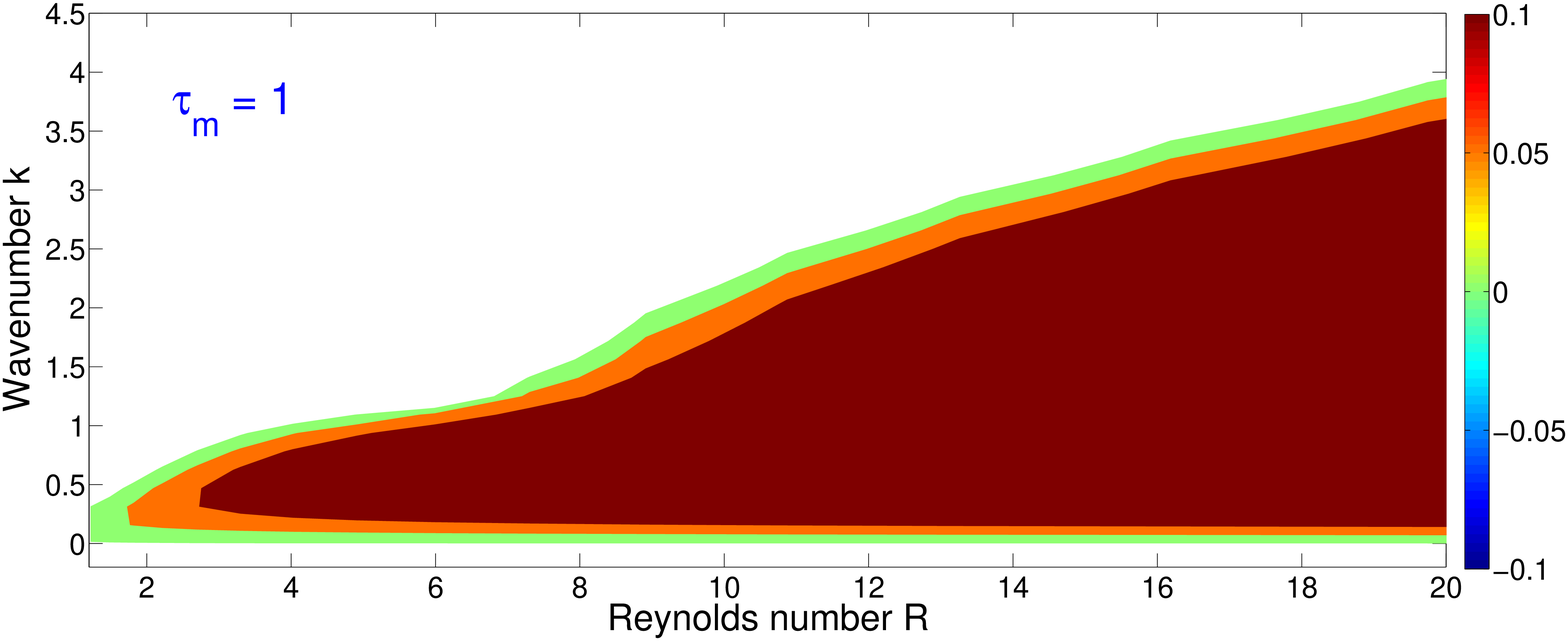}} &
     \resizebox{90mm}{!}{\includegraphics[width=4in,height=3.0in]{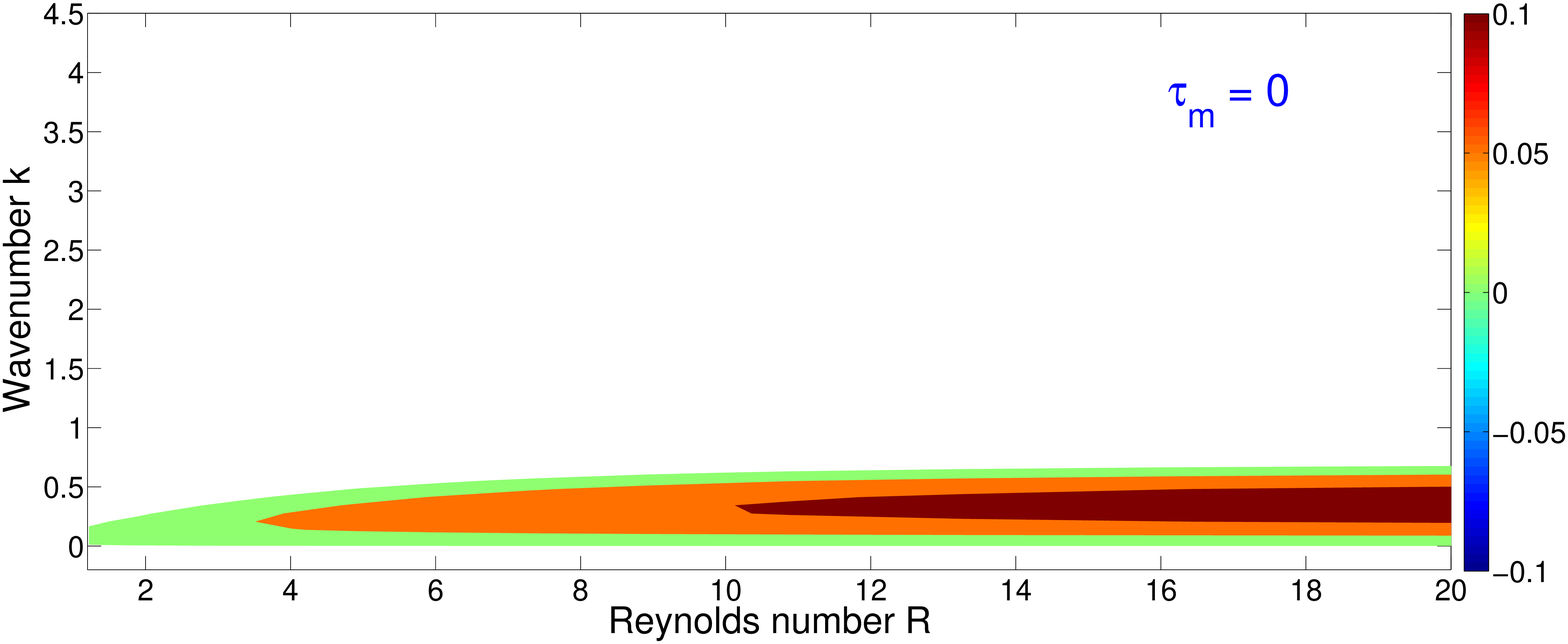}}
     \end{tabular}
     \caption{(Color online)Contour plot of growth rate of KH instability in 2D plane of $k$ and $R$ is shown for  weakly coupled limit $\tau_m = 0$ in the right panel. Colorbar indicates the values of growth rate in different color region. In the left panel, the same contour is plotted for relaxation time $\tau_m = 1$ which shows rapid increase of unstable region in presence of strong coupling.}
     \label{sp1}
   \end{center}
 \end{figure}
\begin{figure}
\centering
\includegraphics[width=5in,height=4.0in]{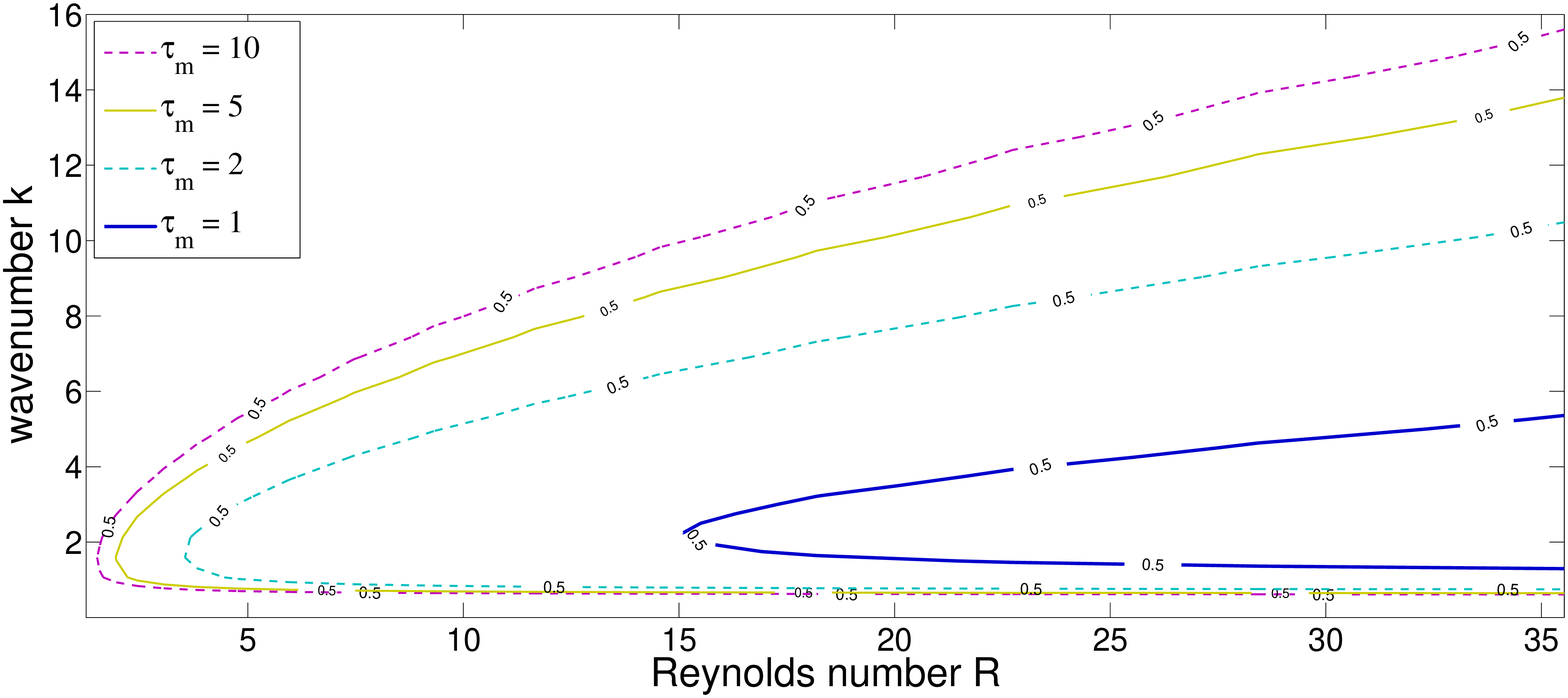}
\caption{(Color online)Contour plot in 2D plane of $k$ and $R$ shows that unstable region increases with increase of relaxation time. Here, curves of growth rate $0.5$ are plotted for different values of $\tau_m$ enlisted in legend.  }
\label{sp2}
\end{figure}

In figure (\ref{sp1}) \& (\ref{sp2}), contour plot is being shown in 2D plane of Reynolds number and wavenumber which clearly shows that unstable region in this parameter space increases with  effect of elasticity.

\section{Summary}
    We have studied Kelvin-Helmholtz instability of dust shear flow  with effects of both viscosity and elasticity in strongly coupled dusty plasma. Viscosity being a dissipative effect, plays stabilizing role. However, elasticity which has energy storing property changes the growth rate of KH instability.
The growth rate of Kelvin-Helmholtz instability   are estimated  using proper Galilean invariant form of the GH equation. The stability characteristics of a small wave-number perturbation are studied analytically by using a discontinuous velocity profile. The combined effect of visco-elastic relaxation time $\tau_m$ and the additional convective term that assures Galilean invariance leads to a modification of the jump conditions and the eigenvalue equation. The results indicate a substantial enhancement of the growth rate and the range of unstable wave numbers over a wide variation of Reynolds number.  The results are further confirmed through a numerical study by choosing a more realistic continuous velocity profile. In the limit $\tau_m \rightarrow 0$, the numerical results reproduce the standard Navier-Stoke's results.
In the absence of the convective term, bunching of the curves is observed with the growth rate vanishing at a particular wave number that is independent of $\tau_m$.
 However, the inclusion of the convective term in the GH operator causes a wide dispersion for the growth rate curves obtained for different values of $\tau_m$ at large values of wave numbers in contrast to the results obtained without the convective term. The results indicate that shear flows are unstable over a large range of wave numbers making their further study useful in context of strongly coupled dusty plasma.

%-----------------------------------------------------------------------
%\bibliography{referenceAPS}% Produces the bibliography via BibTeX.
%------------------------------------------------------------------------
\end{document}